\documentclass[a4paper]{article}

\usepackage{INTERSPEECH2020}
\usepackage{cite}
\usepackage{cancel}
\usepackage[T1]{fontenc}
\usepackage{color}
\usepackage{url}
\usepackage[colorlinks,linkcolor=red,anchorcolor=blue,citecolor=green,CJKbookmarks=True]{hyperref}
\usepackage{multirow}
\usepackage[linesnumbered,ruled,vlined,commentsnumbered]{algorithm2e}

\SetCommentSty{mycommfont}

\newcommand{\pdffigure}[3][width=0.7\linewidth]{
    \begin{figure}[htb]
    \begin{center}
    \IfFileExists{./#2.pdf}{
        \includegraphics[#1]{#2.pdf}
    }{
        \includegraphics[draft]{#2.pdf}
    }
    \end{center}
    \caption{#3}
    \label{fig:#2}
    \end{figure}
}

\title{Speaker-Conditional Chain Model for Speech Separation and Extraction}
\name{Jing Shi$^{1,2}$, 
Jiaming Xu$^1$,Yusuke Fujita$^3$, Shinji Watanabe$^{2}$, Bo Xu$^{1}$}
\address{
  $^1$Institute of Automation, Chinese Academy of Sciences (CASIA)\\
  $^2$Center for Language and Speech Processing, Johns Hopkins University\\
  $^3$Hitachi, Ltd. Research \& Development Group}
\email{shijing2014@ia.ac.cn}

\begin{document}

\maketitle
\begin{abstract}
Speech separation has been extensively explored to tackle the cocktail party problem. However, these studies are still far from having enough generalization capabilities for real scenarios. In this work, we raise a common strategy named Speaker-Conditional Chain Model to process complex speech recordings. In the proposed method, our model first infers the identities of variable numbers of speakers from the observation based on a sequence-to-sequence model. Then, it takes the information from the inferred speakers as conditions to extract their speech sources. With the predicted speaker information from whole observation, our model is helpful to solve the problem of conventional speech separation and speaker extraction for multi-round long recordings. The experiments from standard fully-overlapped speech separation benchmarks show comparable results with prior studies, while our proposed model gets better adaptability for multi-round long recordings. 
\end{abstract}
\noindent\textbf{Index Terms}: speech separation, speaker extraction, cocktail party problem 

\section{Introduction}

Human interactions are often in a broad range of complex auditory scenes, consisting of several speech sources from different speakers and various noises. This complexity poses challenges for many speech technologies, because they usually assume one or zero speaker to be active at the same time~\cite{haeb2019speech}. To tackle these challenging scenes, many techniques have been studied.

Speech separation aims at isolating individual speaker's voices from a recording with overlapped speech~\cite{huang2014deep,wang2016discriminative,hershey2016deep,isik2016single,yu2017permutation,chen2017deep,Drude2018Deep}. With the separation results, both the speech intelligibility for human listening and speech recognition accuracy could be improved~\cite{zeghidour2020wavesplit}. 
Different from the separation task, speaker extraction makes use of additional information to distinguish a target speaker from other participating speakers~\cite{delcroix2018single,wang2019voicefilter,xu2018modeling,xu2020spex}.
Besides, speech denoising~\cite{donahue2018exploring,rethage2018wavenet} and speaker diarization~\cite{FujitaKHNW19,fujita2020end} tasks have also been studied for solving the problem of complex acoustic scenes.

Although many works have been proposed towards each task mentioned above, the processing of natural recordings 
is still challenging. Overall, these tasks are designed to accomplish one particular problem, which has assumptions that do not hold in complex speech recordings.  
For instance, speech separation was heavily explored with pre-segmented audio samples with a length of several seconds (less than 10 seconds), which makes it difficult to form reasonable results for long recordings. Because most existing separation methods only output a fixed number of speech sources with agnostic order, and it is unable to process the variable number of speakers and the relation of the orders between different segments. 
Similarly, the speaker diarization 
bypassed the overlapped part before. Recently, the emergence of EEND approaches~\cite{FujitaKHNW19,fujita2020end} could fix the problem of overlapped speech parts to some extent. However, the diarization results seem an intermediate product without the extraction of each speaker, especially for the overlapped parts.

To address these limitations, we believe that integrating speaker information (used in aim speaker extraction, speaker diarization) into speaker-independent tasks (e.g., speech separation, speech denoising and even speech recognition) will help broaden the application of these techniques towards real scenes.
To be specific, we reconstruct the speech separation/extraction task with the strategy over probabilistic chain rule by importing the conditional probability based on speaker information. In practice, our model automatically infers the information of speakers' identities and then takes it as condition to extract speech sources. 
The speaker information here is some learned hidden representation related to the speaker's identity, which makes it also suitable for open speaker tasks.
We believe this design actually better meets the expectation about an intelligent front-end speech processing pipeline. Because users usually want to get the information about not only the extracted clean speech sources but also which ones speak what.

In this work, we propose our Speaker-Conditional Chain Model (SCCM) to separate the speech sources of different speakers with overlapped speech. Meanwhile, 
the proposed method can handle a long recording with multiple rounds of utterances spoken by different speakers.
Based on this model, we verified its effectiveness in getting both the identity information of each speaker and the extracted speech sources of them. 

The contributions of this paper span the following aspects: (1) we built a common chain model for the processing of speech with one or more speakers. Through the inference-to-extraction pipeline, our model solves the problem about the variable and even unknown number of speakers; (2) with the same architecture, our model shows a comparative performance with the base model, while we could additionally offer accurate speaker identity information for further downstream usage; (3) we proved the effectiveness of this design for both short overlapped segments and long recordings with multi-round conversations, (4) we analyze the advantages and drawbacks of this model.
Our demo video and Supplementary Material are available at \url{https://shincling.github.io/}.

\vspace{-0.2cm}
\section{Related work}
\vspace{-0.1cm}

\subsection{Speech separation}
\vspace{-0.1cm}
As the core part of the cocktail party problem~\cite{cherry1953some}, speech separation gains much attention recently. The common design of this task is to disentangle fully overlapped speech signals from a given short mixture (less than 10 seconds) with a fixed number of speakers. 
Under this design, from spectrogram-based methods~\cite{hershey2016deep,isik2016single,yu2017permutation,Kolbaek2017Multitalker,Luo2018Speaker} to time-domain methods~\cite{luo2018real-time,luo2018tasnet,luo2019dual}, speaker-agnostic separation approaches have been intensively studied.
However, with the steady improvement in performance, most existing approaches might overfit the fully overlapped audio data, which is far from the natural situation with less than 20\% overlap ratio in conversations~\cite{ccetin2006analysis}. Besides, most existing separation models should know the number of speakers in advance and could only tackle the data with the same number of speakers~\cite{shi2018listen}. These constraints further limit their application to real scenes, while our proposed SCCM can provide a solution to the above sparse overlap and unknown speaker number issues. A similar idea with recurrent selective attention networks~\cite{kinoshita2018listening} has been proposed before to tackle the variable number of speakers in separation. However, this model performs with residual spectrograms without leveraging the time-domain methods. And their uPIT~\cite{Kolbaek2017Multitalker} based training is hard to process a long recording, due to the speaker tracing problem raised when chunking the long recording into short segments.


\vspace{-0.2cm}
\subsection{Speaker extraction}
\vspace{-0.1cm}
Another task related to our model is the speaker extraction
~\cite{delcroix2018single,wang2019voicefilter,xu2018modeling,xu2020spex}.
The idea of speaker extraction is to provide a reference from a speaker, and then use such reference to direct the attention to the specified speaker. The reference may be taken from different characteristic binding with the specific speaker, such as voiceprint, location, onset/offset information, and even visual representation~\cite{Ephrat2018Looking}. The speaker extraction technique is particularly useful when the system is expected to respond to a specific target speaker. However, for a meeting or conversation with multiple speakers, the demand for additional references makes it inconvenient. In our work, the reference could be directly inferred from the original recordings, which shows an advantage when the complete analysis of each speaker is needed.


\vspace{-0.2cm}
\section{Speaker-conditional chain model}

\begin{figure}[t]
  \centering
  \includegraphics[width=\linewidth]{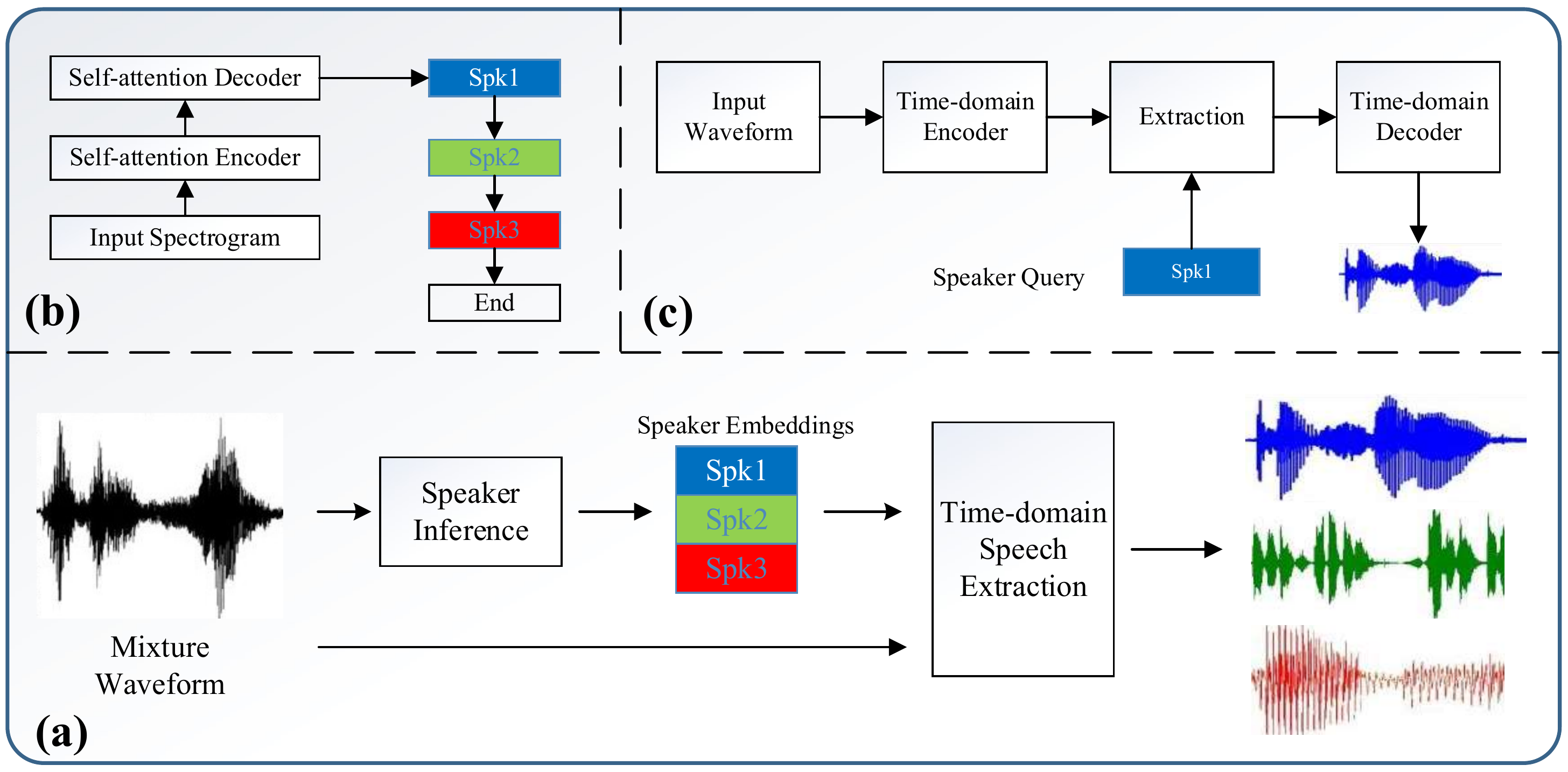}
  \caption{The framework of the proposed Speaker-Conditional Chain Model (SCCM). (a) shows the whole strategy of our proposed SCCM; (b) is the module of \textbf{speaker inference}, which predicts the speaker identities and corresponding embeddings. 
  (c) refers to the time-domain \textbf{speech extraction} module. This module takes the each inferred information from (b) respectively to conduct a conditional extraction.}
  \label{fig:framework}
\vspace{-0.5cm}
\end{figure}

This section describes our Speaker-Conditional Chain Model (SCCM). 
As illustrated in Figure~\ref{fig:framework}, the chain here refers to a pipeline through two sequential components: speaker inference and speech extraction.
These models are integrated based on a joint probability formulation, which will be described in Section \ref{sec:formulation}.
Speaker identities play an important role in our strategy. 
The speaker inference module aims to predict the possible speaker identities and the corresponding embedding vectors.
The speech extraction module takes each embedding from the speaker inference module as the query to disentangle the corresponding source audio from the input recording.

This design will bring several advantages. 
First, the possible speakers are inferred by a sequence-to-sequence model with an end-of-sequence label, which easily handles variable and unknown numbers of speakers. 
Second, the inference part is based on a self-attention network, which utilizes the full context information in a recording to form a speaker embedding. 
This avoids the calculation inefficiency problem in some clustering-based models~\cite{zeghidour2020wavesplit,hershey2016deep,isik2016single}, which needs an iterative k-means algorithm in each frame. 
Third, the information about each speaker will make it suitable for our model to some further applications in speaker diarization or speaker tracking.

\vspace{-0.2cm}
\subsection{Problem setting and formulation}
\label{sec:formulation}
Assume there is a training dataset with a set of speaker identities $\mathcal{Y}$ with $|\mathcal{Y}|=N$ known distinct speakers in total.
In a $T$-length segment of waveform observation $O\in \mathbb{R}^T$, there are $I$ different speakers $Y=(y_1,...,y_i,...,y_I)$.
Each speaker $y_i$\footnote{Although $y_i \in \mathcal{Y}$ during training, potentially $y_i \notin \mathcal{Y}$  during inference in the open speaker task, where the system could still provide a meaningful speaker embedding vector for downstream applications.} has the corresponding speech source $s_i\in \mathbb{R}^T$ to form the set of sources $S=(s_1,...,s_i,...,s_I)$. 
The basic formulation of our strategy is to estimate the joint probability of speaker labels and corresponding sources, i.e., $p(S, Y|O)$.
This is factorized with speaker inference probability $p(Y|O)$ and speech extraction probability $p(S|Y, O)$ as follows:
\begin{equation}
p(S,Y|O) = p(S|Y,O)p(Y|O).
  \label{eq1}
\end{equation}
We further factorize each probability distribution based on the probabilistic chain rule.

The speaker inference probability $p(Y|O)$ in Eq.~\eqref{eq1} recursively predicts variable numbers of speaker identities as follows:
\begin{equation}
  p(Y|O)= \prod_i p(y_i|O,y_{i-1},...,y_{1}).
  \label{eq2}
\end{equation}
We adopt a \textit{sequence-to-sequence model} based on self-attention transformer~\cite{vaswani2017attention}, as illustrated in Figure~\ref{fig:framework}(b).
The network architecture of $p(Y|O)$ will be discussed in Section \ref{sec:speaker_inference}.

The speech extraction probability $p(S|Y,O)$ in Eq.~\eqref{eq1} is also factorized by using the probabilistic chain rule and the conditional independence assumption, as follows:
\begin{equation}
  p(S|Y,O)= \prod_i p(s_i|y_i, \cancel{y_{\setminus i}, s_{1},..., s_{i-1}} \,O) = \prod_i p(s_i|y_{i},O).
  \label{eq3}
\end{equation}
As illustrated in Figure~\ref{fig:framework}(c), our speech extraction module takes the speaker identity $y_i$, which is predicted from the speaker inference module $p(Y|O)$ in Eq.~\eqref{eq2}, to conduct a conditional extraction.
Every speaker information here serves as the condition to guide the following extraction. 
For multi-round long recordings, the speaker information will be formed as global information from the whole observation to track the specific speaker. 
The network architecture of $p(s_i|y_i, O)$ will be discussed in Section \ref{sec:speech_extraction}.


\subsection{Speaker inference module}
\label{sec:speaker_inference}
In the speaker inference part, we build a model to simulate the probability $p(Y|O)$ in Eq.~(\ref{eq1}) and (\ref{eq2}) .
We adopt a self-attention based transformer \cite{vaswani2017attention} architecture as the encoder-decoder structure.
In this part, we take the observation spectrogram (Short-Time Fourier Transform (STFT) coefficients) as an input. 
The reason we do not use the time-domain approach here is to avoid excessive computation complexities which may consume too much GPU memory to train the model, especially with inputs of long recordings.

In detail, for a given spectrogram $\mathbf{X}$ containing $\tilde{T}$ frames and $F$ frequency bins, it is viewed as a sequence of frames. 
For the encoder part, we use the Transformer Encoder as follows:
\begin{align}
    \mathbf{E}_0 \,\, &= \text{Linear}^{(F \mapsto D)}(\mathbf{X}) \in \mathbb{R}^{D \times \tilde{T}}, \\
    \mathbf{E}_m &= \text{Encoder}(\mathbf{E}_{m-1}) \in \mathbb{R}^{D \times \tilde{T}} & (1 \le m \le M),
\end{align}
where, $\text{Linear}^{(F \mapsto D)}()$ is a linear projection that maps $F$-dimensional vector to $D$-dimensional vector for each column of the input matrix.
$\text{Encoder}()$ is the Transformer Encoder block that contains multi-head self-attention layer, position-wise feed-forward layer, and residual connections. By stacking the encoder $M$ times, $\mathbf{E}_M \in \mathbb{R}^{D \times \tilde{T}}$ is an output of the encoder part.

For the decoder part, the neural network outputs probability distribution $\mathbf{z}_i$ for the $i$-th speaker, calculated as follows:
\begin{align}
    \label{eq:pos}
    \mathbf{j}_i\, &= \mathrm{Linear}^{(1 \mapsto D)}(\mathbf{i}) \in \mathbb{R}^{D},\\ 
    \mathbf{h}_i &= \mathrm{Decoder}(\mathbf{E}_M, \mathbf{h}_{i-1},\mathbf{j}_i) \in \mathbb{R}^{D}, \label{eq:embedding}\\
    \mathbf{z}_i &= \mathrm{Softmax}(\mathrm{Linear}^{(D \mapsto \tilde{N})}(\mathbf{h}_i)) \in \mathbb{R}^{N+1},
\end{align}
where $\mathbf{j}_i$ is the positional encoding in each step to predict the speaker. 
$\text{Decoder}()$ is the Transformer Decoder block, which takes the states from the output of encoder and the hidden state from the previous step to output the speakers embedding $\mathbf{h}_{i}$ at this step. 
Finally, a linear projection with a softmax produces a $(N+1)$-dimensional vector $\mathbf{z}_i$ as the network output, where $\mathbf{z}_i$ is the $i$-th predicted probability distribution over the union of speaker set $\mathcal{Y}$ and the additional end-of-sequence label $\langle\mathrm{EOS}\rangle$, i.e., $y^* \in \{\mathcal{Y}, \langle\mathrm{EOS}\rangle\}$.

\subsection{Speech extraction module}
\label{sec:speech_extraction}
\vspace{-0.1cm}
For the speech extraction module, each speaker channel $p(s_i|y_i, O)$ will be processed independently, as formed in Eq.~(\ref{eq3}).
This part takes each inferred speaker embedding $\mathbf{h}_i$ predicted in Eq.~\eqref{eq:embedding} instead of identity $y_i$, and the raw waveform $O$ as input to produce the corresponding clean signal $\mathbf{\hat{s}_i}$:
\vspace{-0.15cm}
\begin{align}
    \mathbf{\hat{s}_i} &= \mathrm{Extractor}(O, \mathbf{h}_{i}) \in \mathbb{R}^{T},
\end{align}
where, $\text{Extractor}()$ takes a similar architecture with time-domain speech separation methods from the Conv-TasNet~\cite{luo2018tasnet}. 
The difference lies in that we will output one channel towards each speaker embedding rather than separate several sources together. 
To be specific, at the end of the separator module in~\cite{luo2018tasnet}, we will concatenate the $\mathbf{h}_{i}$ with each frame of the output features. Then, a single channel $1\times1-conv$ operation is conducted towards this speaker, rather than multi-channel (as the number of speakers in this mixture).  
Besides this simple fusion approach, we have tested several different methods to integrate the condition vector $\mathbf{h}_{i}$ into the model. For example, to concatenate it at the beginning of the separator, or use the similar method in~\cite{zeghidour2020wavesplit} with FiLM~\cite{perez2018film} in each block in TasNet's separator. However, we found both of the other methods cause severe overfitting.

\subsection{Training targets}
Our whole model is end-to-end, with the loss $\mathcal{L}$, which corresponds to optimize the joint probability $p(S, Y|O)$ in Eq.~\eqref{eq1}.
$\mathcal{L}$ is calculated from both the cross-entropy loss $\mathcal{L}_c$, which corresponds to deal with speaker inference $p(Y|O)$ in Section \ref{sec:speaker_inference}, and the source reconstruction loss (SI-SNR) $\mathcal{L}_r$, which corresponds to deal with speech extraction $p(S|Y, O)$ in a non-probabilistic manner in Section \ref{sec:speech_extraction}. 
One critical problem in training SCCM is to decide the order of the inferred speakers. For one possible permutation $\rho$, the speakers list $Y$ and the speech sources $S$ will be re-ordered synchronously as follows: 
\vspace{-0.2cm}
\begin{align}
     Y_\rho&=(y_1^*,y_2^*...,y_I^*), \forall{\ y}_i^*\stackrel{\rho}\Rightarrow y \in Y,\\
     S_\rho&=(s_1^*,s_2^*...,s_I^*), \,\forall{\ s}_i^*\stackrel{\rho}\Rightarrow \, s \in S .
\end{align}
Some former works have shown that the seq2seq structure helps to improve the accuracy in the inference module by setting a fixed order in training~\cite{shi2019ones}. We compared several options to use a random fixed order or use the order defined by the energy in the spectrogram (observed well in~\cite{weng2015deep}). But we found the order decided by the model itself gets better performance in practice. Therefore, we take the best permutation $\theta$ with least reconstruction error in the extraction part as the order to train the inference part as follows:
\vspace{-0.2cm}
\begin{align}
		\theta &= \mathop{\mathrm{argmin}}_{\mathrm{\rho\sim Perms}}{\mathcal{L}_r({\mathbf{\hat{S}}},S_\rho)}, \\
     \mathcal{L}&=\mathcal{L}_r({\mathbf{\hat{S}},S_\theta)} + \alpha \,\mathcal{L}_c({\mathbf{Z},Y_\theta}),\label{eq:loss}
\end{align}
where we use $\alpha=50$ in all our experiments.

\vspace{-0.1cm}
\section{Experiments}
\vspace{-0.1cm}
As a generalized framework to tackle the problem of extracting speech sources of all speakers, we tested the effectiveness of SCCM with different tasks. Besides the signal reconstruction quality (e.g., SDRi, SI-SNRi) used in speech separation task, we also verified the performance over speaker identification and speech recognition. 
In our experiments, all data are resampled to 8 kHz. For the speaker inference module, the magnitude spectra are used as the input feature, computed from STFT with 32 ms window length, 8 ms hop size, and the sine window.
More detailed configuration of the proposed architecture could be seen in Section A.1 of our \href{https://drive.google.com/file/d/1aqJy465dLHaWPdMqG-BgjAgYEg70q7as/view?usp=sharing}{Supplementary Material}\footnote{\url{https://drive.google.com/open?id=1aqJy465dLHaWPdMqG-BgjAgYEg70q7as}}
.

\vspace{-0.2cm}
\subsection{Speech separation for overlapped speech}
First, we evaluated our method on fully-overlapped speech mixtures from the Wall Street Journal (WSJ0) corpus. The WSJ0-2mix and 3mix datasets are the benchmarks designed for speech separation in \cite{hershey2016deep}. 
In the validation set, we used the so-called Closed Conditions (CC) in~\cite{hershey2016deep,isik2016single}, where the speakers are all from the training set. 
As a contrast, for the evaluation set, we use Open Condition (OC), which provides unknown speakers. 
For the separation performance, we compare our results with the TasNet, which is our base model described in Section ~\ref{sec:speech_extraction}, without changing any hyper-parameter.  
Table~\ref{tab:ss} listed the speech separation performance over the different training sets. 

Table~\ref{tab:ss} shows that our SCCM got slightly worse performance than the base model in OC with the same architecture and training dataset.
However, unlike the fixed-speaker-number speech separation method, SCCM could be trained and tested in the variable number of speakers with a single model thanks to our speaker-conditional strategy with the sequence-to-sequence model.
As we expect, the training with both WSJ0-2mix and WSJ0-3mix datasets got better performance than the training with each dataset in close condition.
Although we did not achieve obvious improvement in the OC case, with the careful tuning based on the cascading technique (the similar methods used in~\cite{Kolbaek2017Multitalker}), the separation performance gets a notable improvement, which also exceeds the base model. 
For the SCCM+ model, we use the extracted speech source, along with the raw observation, as input to go through another extraction module (TasNet). With this cascading method, the details of the extracted source get further optimized, which may fix the ambiguity caused by the independence assumption in Eq. (\ref{eq3}).    

Also, as the former node in the chain, the ability to predict the correct speakers or get the distinct and informative embeddings is quite crucial.  Table~\ref{tab:si} shows the performance of the speaker inference module, as discussed in Section \ref{sec:speaker_inference}. 
For the CC, micro-F1 is calculated to evaluate the correctness of the predicted speakers. For the OC, we use the speaker counting accuracy to measure the speaker inference module, which guarantees the success of the subsequent speech extraction module. 
From the results, we could see that the speaker inference module in SCCM could reasonably infer the correct speaker identity in CC and the correct number of speakers in OC.

It should be mentioned that the number of speakers in training data ($N$ in Section \ref{sec:formulation})  with WSJ0-2mix and 3mix is 101, much smaller than the number in a standard speaker recognition task (e.g., 1,211 in VoxCeleb1~\cite{Nagrani17}). We infer that this limited number somewhat limits the performance of the speaker inference part and the following extraction module, especially for the open condition. Besides, compared with the state-of-the-art speaker recognition methods, our model takes the overlapped speech as input, which also brings more complexity.

\begin{table}[tb]
\vspace{-0.3cm}
  \caption{Speech separation performance (SI-SNRi) for the benchmark datasets with overlapped speech.}
\vspace{-0.3cm}
  \label{tab:ss}
  \centering
\resizebox{8cm}{1.15cm}{
\begin{tabular}{c|c|c|c|c|c}
\hline\hline
\multirow{2}{*}{Models} & \multirow{2}{*}{\begin{tabular}[c]{@{}c@{}}Training\\ Dataset\end{tabular}} & \multicolumn{2}{c|}{SI-SNRi CC} & \multicolumn{2}{c}{SI-SNRi OC} \\ \cline{3-6} 
 &  & WSJ0-2mix & WSJ0-3mix & WSJ0-2mix & WSJ0-3mix \\ \hline
\multirow{2}{*}{TasNet} & WSJ0-2mix & - & - & 14.6 & - \\ \cline{2-6} 
 & WSJ0-3mix & - & - & - & 11.6 \\ \hline
\multirow{3}{*}{SCCM} & WSJ0-2mix & 15.4 & - & 14.5 & - \\ \cline{2-6} 
 & WSJ0-3mix & - & 11.9 & - & 11.4 \\ \cline{2-6} 
 & both & 16.4 & 12.1 & 14.3 & 11.3 \\ \hline
SCCM+ & both & \bf{17.7} & \bf{13.4} & \textbf{15.4} & \textbf{12.5} \\ \hline\hline
\end{tabular}}
\end{table}
\vspace{-0.1cm}

\begin{table}[t]
\vspace{-0.2cm}
  \caption{Speaker inference performance of SCCM.}
\vspace{-0.3cm}
  \label{tab:si}
  \centering
\resizebox{7cm}{0.9cm}{
\begin{tabular}{c|c|c|c|c}
\hline\hline
\multirow{2}{*}{\begin{tabular}[c]{@{}c@{}}Training\\ Dataset\end{tabular}} & \multicolumn{2}{c|}{\begin{tabular}[c]{@{}c@{}}F1 scores in\\ Validset (CC)\end{tabular}} & \multicolumn{2}{c}{\begin{tabular}[c]{@{}c@{}}Speaker counting accuracy in\\ Testset (OC)\end{tabular}} \\ \cline{2-5} 
 & WSJ0-2mix & WSJ0-3mix & WSJ0-2mix & WSJ0-3mix \\ \hline
WSJ0-2mix & 89.2 &- & \bf{99.7} & - \\ \hline
\multicolumn{1}{l|}{WSJ0-3mix}  & - & \multicolumn{1}{c|}{-} & - & \bf{98.9} \\ \hline
both & \bf{90.4}&  \bf{75.5} & 96.8 & 94.5 \\ \hline\hline
\end{tabular}}
\vspace{-0.5cm}
\end{table}
\vspace{-0.1cm}

\subsection{Extraction performance for multi-round recordings}\label{sec:multi-round}
\vspace{-0.1cm}
As mentioned before, the natural conversions in real scenes usually get multi-round utterances from several speakers. And the ratio of overlapped speech is less than 20\% in general. For the conventional speech separation methods, there exists a problem with the consistent order of several speakers in different parts in a relatively long recording, especially when the dominant speaker changes~\cite{zeghidour2020wavesplit}. To validate this, we extend each mixture in the standard WSJ0-mix to multiple rounds. 
In detail (seen in Algorithm 1 and Section A.2 in \href{https://drive.google.com/file/d/1aqJy465dLHaWPdMqG-BgjAgYEg70q7as/view?usp=sharing}{Supplementary Material}), we take the list of the original mixtures from WSJ0-2mix and sample several additional utterances from the provided speakers. 
After getting the sources from different speakers, the long recording will be formed by concatenating the sources one by one. The beginning of the following source gets a random shift around the end of the former one, making it similar to a natural conversation with an overlap-ratio around 15\%.

Without any change in our model, we could directly train our SCCM on the synthetic multi-round data. 
It should be mentioned that our speaker inference module takes the whole spectrogram as an input. 
In contrast, the speech extraction module takes a random segment with 4 seconds from the long recording to avoid the problem with out-of-memory. 
Table~\ref{tab:multi} shows the performance difference compared with the base model. 
Both valid set and test set are fixed with four rounds of conversations with an average length of 10 seconds. As we expect, the results show that SCCM stays more stable than the baseline model with multi-round recordings. To further understand the model, we observed the attention status of the Decoder in Eq.~(\ref{eq:embedding}). 
We find the attention of the inference reflects the speaker's activities at different parts within a recording. 
More details and visualization could be viewed in Section A.3 in the \href{https://drive.google.com/file/d/1aqJy465dLHaWPdMqG-BgjAgYEg70q7as/view?usp=sharing}{Supplementary Material}.

\begin{table}[t]
\vspace{-0.3cm}
  \caption{Extraction performance for multi-round recordings.}
\vspace{-0.3cm}
  \label{tab:multi}
  \centering
\resizebox{4.5cm}{0.9cm}{
\begin{tabular}{c|c|c|c|c}
\hline\hline
       & \multicolumn{2}{c|}{Valid SI-SNRi}    & \multicolumn{2}{c}{Test SI-SNRi}      \\ \hline
TasNet & \multicolumn{2}{c|}{14.2}          & \multicolumn{2}{c}{11.5}            \\ \hline
SCCM   & \multicolumn{2}{c|}{\textbf{17.5}} & \multicolumn{2}{c}{\textbf{13.7}}           \\ \hline
       & \textless{}5dB & \textgreater{}5dB & \textless{}5dB & \textgreater{}5dB \\ \hline
TasNet & 17.0\%           & 83.0\%              & 33.6\%           & 66.4\%               \\ \hline
SCCM   & 12.6\%           & 87.4\%              & 26.8\%           & 73.2\%               \\ \hline\hline
\end{tabular}}
\vspace{-0.3cm}
\end{table}

\vspace{-0.1cm}
\begin{table}[t]
  \caption{WERs for utterance-wise evaluation over the single-channel LibriCSS dataset with clean mixtures. 0S: 0\% overlap with short inter-utterance silence (0.1-0.5 s). 0L: 0\% overlap with long inter-utterance silence (2.9-3.0 s).}
\vspace{-0.3cm}
  \label{tab:css}
  \centering
\resizebox{7cm}{0.7cm}{
\begin{tabular}{c|cccccc}
\hline\hline
\multirow{2}{*}{System} & \multicolumn{6}{c}{Overlap ratio in \%} \\
 & 0S & 0L & 10 & 20 & 30 & 40 \\ \hline
No separation & \multicolumn{1}{c|}{2.7} & \multicolumn{1}{c|}{3.0} & \multicolumn{1}{c|}{11.9} & \multicolumn{1}{c|}{20.4} & \multicolumn{1}{c|}{30.2} & 43.0 \\ \hline
Single-channel SCCM & \multicolumn{1}{c|}{9.5} & \multicolumn{1}{c|}{9.4} & \multicolumn{1}{c|}{6.5} & \multicolumn{1}{c|}{9.3} & \multicolumn{1}{c|}{11.9} & 13.9 \\ \hline\hline
\end{tabular}}
\vspace{-0.4cm}
\end{table}
\vspace{-0.0cm}

\subsection{Speech recognition in continuous speech separation}
To further validate the downstream application, we conducted the speech recognition in the recently proposed continuous speech separation dataset~\cite{chen2020continuous}. LibriCSS is derived from LibriSpeech~\cite{panayotov2015librispeech} by concatenating the corpus utterances to simulate conversations. In line with the utterance-wise evaluation in LibriCSS, we directly use our trained model from the former multi-round task to test the recognition performance. The original raw recordings in LibriCSS are from far-field scenes with noise and reverberation, which is inconsistent with ours. So we use the single-channel clean mixtures and convert to 8 kHz to separate them. Moreover, we use the trained model from the Espnet's~\cite{watanabe2018espnet} LibriSpeech recipe to recognize each utterance. Table~\ref{tab:css} shows the WER results in this dataset.  

We observed that (1) the results show a similar tendency with the provided baseline model in LibriCSS~\cite{chen2020continuous}. (2) With the increase of overlap ratio, the performance on the original clean mixture  becomes much worse, while our model stays a low level of WER. (3) Because the training data of our model comes from the situation of multi speakers, the performance on the no-overlapped segments becomes worse. And we think this could be avoided by adding some single speaker's segments in the training set.

\vspace{-0.2cm}
\section{Conclusions}
\vspace{-0.1cm}
We introduced the Speaker-conditional chain model as a common framework to process audio recordings with multiple speakers. Our model could be applied to tackle the separation problem towards fully-overlapped speech with variable and unknown number of speakers. Meanwhile,  multi-round long audio recordings in natural scenes can also be modeled and extracted effectively using this method.  Experimental results showed the effectiveness and good adaptability of the proposed model. Our following work will extend this model to the real scenes with noisy and reverberant multi-channel recordings. We would also like to explore the factors to improve the generalization ability of this approach, like the introduction of more speakers or changes in the network and training objectives.

\newpage
\bibliographystyle{IEEEtran}

\bibliography{interspeech2020}

\begin{thebibliography}{10}
\providecommand{\url}[1]{#1}
\csname url@samestyle\endcsname
\providecommand{\newblock}{\relax}
\providecommand{\bibinfo}[2]{#2}
\providecommand{\BIBentrySTDinterwordspacing}{\spaceskip=0pt\relax}
\providecommand{\BIBentryALTinterwordstretchfactor}{4}
\providecommand{\BIBentryALTinterwordspacing}{\spaceskip=\fontdimen2\font plus
\BIBentryALTinterwordstretchfactor\fontdimen3\font minus
  \fontdimen4\font\relax}
\providecommand{\BIBforeignlanguage}[2]{{%
\expandafter\ifx\csname l@#1\endcsname\relax
\typeout{** WARNING: IEEEtran.bst: No hyphenation pattern has been}%
\typeout{** loaded for the language `#1'. Using the pattern for}%
\typeout{** the default language instead.}%
\else
\language=\csname l@#1\endcsname
\fi
#2}}
\providecommand{\BIBdecl}{\relax}
\BIBdecl

\bibitem{haeb2019speech}
R.~Haeb-Umbach, S.~Watanabe, T.~Nakatani, M.~Bacchiani, B.~Hoffmeister, M.~L.
  Seltzer, H.~Zen, and M.~Souden, ``Speech processing for digital home
  assistants: Combining signal processing with deep-learning techniques,''
  \emph{IEEE Signal Processing Magazine}, vol.~36, no.~6, pp. 111--124, 2019.

\bibitem{huang2014deep}
P.~Huang, M.~Kim, M.~Hasegawajohnson, and P.~Smaragdis, ``Deep learning for
  monaural speech separation,'' in \emph{ICASSP}, 2014, pp. 1562--1566.

\bibitem{wang2016discriminative}
G.~Wang, C.~Hsu, and J.~Chien, ``Discriminative deep recurrent neural networks
  for monaural speech separation,'' in \emph{ICASSP}, 2016, pp. 2544--2548.

\bibitem{hershey2016deep}
J.~R. Hershey, Z.~Chen, J.~Le~Roux, and S.~Watanabe, ``Deep clustering:
  Discriminative embeddings for segmentation and separation,'' in
  \emph{ICASSP}, 2016, pp. 31--35.

\bibitem{isik2016single}
Y.~Isik, J.~L. Roux, Z.~Chen, S.~Watanabe, and J.~R. Hershey, ``Single-channel
  multi-speaker separation using deep clustering,'' in \emph{INTERSPEECH},
  2016.

\bibitem{yu2017permutation}
D.~Yu, M.~Kolb{\ae}k, Z.-H. Tan, and J.~Jensen, ``Permutation invariant
  training of deep models for speaker-independent multi-talker speech
  separation,'' in \emph{ICASSP}, 2017, pp. 241--245.

\bibitem{chen2017deep}
Z.~Chen, Y.~Luo, and N.~Mesgarani, ``Deep attractor network for
  single-microphone speaker separation,'' in \emph{ICASSP}, 2017, pp. 246--250.

\bibitem{Drude2018Deep}
L.~Drude, T.~von Neumann, and R.~Haeb-Umbach, ``Deep attractor networks for
  speaker re-identification and blind source separation,'' in \emph{ICASSP},
  2018, pp. 11--15.

\bibitem{zeghidour2020wavesplit}
N.~Zeghidour and D.~Grangier, ``Wavesplit: End-to-end speech separation by
  speaker clustering,'' \emph{arXiv preprint arXiv:2002.08933}, 2020.

\bibitem{delcroix2018single}
M.~{Delcroix}, K.~{Zmolikova}, K.~{Kinoshita}, A.~{Ogawa}, and T.~{Nakatani},
  ``Single channel target speaker extraction and recognition with speaker
  beam,'' in \emph{ICASSP}, 2018, pp. 5554--5558.

\bibitem{wang2019voicefilter}
Q.~Wang, H.~Muckenhirn, K.~Wilson, P.~Sridhar, Z.~Wu, J.~R. Hershey, R.~A.
  Saurous, R.~J. Weiss, Y.~Jia, and I.~L. Moreno, ``{VoiceFilter: Targeted
  Voice Separation by Speaker-Conditioned Spectrogram Masking},'' in
  \emph{INTERSPEECH}, 2019, pp. 2728--2732.

\bibitem{xu2018modeling}
J.~Xu, J.~Shi, G.~Liu, X.~Chen, and B.~Xu, ``Modeling attention and memory for
  auditory selection in a cocktail party environment,'' in \emph{Proceedings of
  the 32nd AAAI Conference on Artificial Intelligence (AAAI)}, 2018, pp.
  2564--2571.

\bibitem{xu2020spex}
C.~Xu, W.~Rao, E.~S. Chng, and H.~Li, ``Spex: Multi-scale time domain speaker
  extraction network,'' \emph{arXiv preprint arXiv:2004.08326}, 2020.

\bibitem{donahue2018exploring}
C.~Donahue, B.~Li, and R.~Prabhavalkar, ``Exploring speech enhancement with
  generative adversarial networks for robust speech recognition,'' in
  \emph{ICASSP}, 2018, pp. 5024--5028.

\bibitem{rethage2018wavenet}
D.~Rethage, J.~Pons, and X.~Serra, ``A wavenet for speech denoising,'' in
  \emph{ICASSP}, 2018, pp. 5069--5073.

\bibitem{FujitaKHNW19}
Y.~Fujita, N.~Kanda, S.~Horiguchi, K.~Nagamatsu, and S.~Watanabe, ``End-to-end
  neural speaker diarization with permutation-free objectives,'' in
  \emph{INTERSPEECH}, 2019, pp. 4300--4304.

\bibitem{fujita2020end}
Y.~Fujita, N.~Kanda, S.~Horiguchi, Y.~Xue, K.~Nagamatsu, and S.~Watanabe,
  ``End-to-end neural speaker diarization with self-attention,'' in
  \emph{ASRU}, 2019.

\bibitem{cherry1953some}
E.~C. Cherry, ``Some experiments on the recognition of speech, with one and
  with two ears,'' \emph{Journal of the Acoustical Society of America},
  vol.~25, no.~5, pp. 975--979, 1953.

\bibitem{Kolbaek2017Multitalker}
M.~Kolbaek, D.~Yu, Z.~H. Tan, J.~Jensen, M.~Kolbaek, D.~Yu, Z.~H. Tan, and
  J.~Jensen, ``Multitalker speech separation with utterance-level permutation
  invariant training of deep recurrent neural networks,'' \emph{IEEE/ACM
  Transactions on Audio Speech and Language Processing}, vol.~25, no.~10, pp.
  1901--1913, 2017.

\bibitem{Luo2018Speaker}
Y.~Luo, Z.~Chen, and N.~Mesgarani, ``Speaker-independent speech separation with
  deep attractor network,'' \emph{IEEE/ACM Transactions on Audio Speech and
  Language Processing}, vol.~26, no.~4, pp. 787--796, 2018.

\bibitem{luo2018real-time}
Y.~Luo and N.~Mesgarani, ``Real-time single-channel dereverberation and
  separation with time-domain audio separation network.'' in
  \emph{INTERSPEECH}, 2018, pp. 342--346.

\bibitem{luo2018tasnet}
------, ``Tasnet:time-domain audio separation network for real-time,
  single-channel speech separation,'' in \emph{ICASSP}, 2018, pp. 696--700.

\bibitem{luo2019dual}
Y.~Luo, Z.~Chen, and T.~Yoshioka, ``Dual-path rnn: efficient long sequence
  modeling for time-domain single-channel speech separation,'' in
  \emph{ICASSP}, 2020, pp. 46--50.

\bibitem{ccetin2006analysis}
{\"O}.~{\c{C}}etin and E.~Shriberg, ``Analysis of overlaps in meetings by
  dialog factors, hot spots, speakers, and collection site: Insights for
  automatic speech recognition,'' in \emph{Ninth international conference on
  spoken language processing}, 2006.

\bibitem{shi2018listen}
J.~Shi, J.~Xu, G.~Liu, and B.~Xu, ``Listen, think and listen again: Capturing
  top-down auditory attention for speaker-independent speech separation,'' in
  \emph{Proceedings of the 27th International Joint Conference on Artificial
  Intelligence (IJCAI)}, 2018.

\bibitem{kinoshita2018listening}
K.~Kinoshita, L.~Drude, M.~Delcroix, and T.~Nakatani, ``Listening to each
  speaker one by one with recurrent selective hearing networks,'' in
  \emph{ICASSP}, 2018, pp. 5064--5068.

\bibitem{Ephrat2018Looking}
A.~Ephrat, I.~Mosseri, O.~Lang, T.~Dekel, and M.~Rubinstein, ``Looking to
  listen at the cocktail party: A speaker-independent audio-visual model for
  speech separation,'' \emph{Acm Transactions on Graphics}, vol.~37, no.~4, pp.
  1--11, 2018.

\bibitem{vaswani2017attention}
A.~Vaswani, N.~Shazeer, N.~Parmar, J.~Uszkoreit, L.~Jones, A.~N. Gomez,
  {\L}.~Kaiser, and I.~Polosukhin, ``Attention is all you need,'' in
  \emph{Advances in neural information processing systems}, 2017, pp.
  5998--6008.

\bibitem{perez2018film}
E.~Perez, F.~Strub, H.~De~Vries, V.~Dumoulin, and A.~Courville, ``Film: Visual
  reasoning with a general conditioning layer,'' in \emph{Thirty-Second AAAI
  Conference on Artificial Intelligence}, 2018.

\bibitem{shi2019ones}
J.~Shi, J.~Xu, and B.~Xu, ``Which ones are speaking? speaker-inferred model for
  multi-talker speech separation,'' \emph{INTERSPEECH}, pp. 4609--4613, 2019.

\bibitem{weng2015deep}
C.~Weng, D.~Yu, M.~L. Seltzer, and J.~Droppo, ``Deep neural networks for
  single-channel multi-talker speech recognition,'' \emph{IEEE/ACM Transactions
  on Audio, Speech, and Language Processing}, vol.~23, no.~10, pp. 1670--1679,
  2015.

\bibitem{Nagrani17}
A.~Nagrani, J.~S. Chung, and A.~Zisserman, ``Voxceleb: a large-scale speaker
  identification dataset,'' in \emph{INTERSPEECH}, 2017.

\bibitem{chen2020continuous}
Z.~Chen, T.~Yoshioka, L.~Lu, T.~Zhou, Z.~Meng, Y.~Luo, J.~Wu, and J.~Li,
  ``Continuous speech separation: dataset and analysis,'' \emph{arXiv preprint
  arXiv:2001.11482}, 2020.

\bibitem{panayotov2015librispeech}
V.~Panayotov, G.~Chen, D.~Povey, and S.~Khudanpur, ``Librispeech: an asr corpus
  based on public domain audio books,'' in \emph{ICASSP}, 2015, pp. 5206--5210.

\bibitem{watanabe2018espnet}
S.~Watanabe, T.~Hori, S.~Karita, T.~Hayashi, J.~Nishitoba, Y.~Unno, N.~E.~Y.
  Soplin, J.~Heymann, M.~Wiesner, N.~Chen \emph{et~al.}, ``Espnet: End-to-end
  speech processing toolkit,'' in \emph{INTERSPEECH}, 2018.

\bibitem{luo2019conv}
Y.~Luo and N.~Mesgarani, ``Conv-tasnet: Surpassing ideal time--frequency
  magnitude masking for speech separation,'' \emph{IEEE/ACM transactions on
  audio, speech, and language processing}, vol.~27, no.~8, pp. 1256--1266,
  2019.

\bibitem{bahdanau2014neural}
D.~Bahdanau, K.~Cho, and Y.~Bengio, ``Neural machine translation by jointly
  learning to align and translate,'' \emph{Computer Science}, 2014.

\bibitem{kim2017structured}
Y.~Kim, C.~Denton, L.~Hoang, and A.~M. Rush, ``Structured attention networks,''
  in \emph{5th International Conference on Learning Representations {ICLR},
  Toulon, France, April 24-26}, 2017.

\end{thebibliography}
\newpage
\begin{appendix}

\section{Supplementary Material}
\begin{algorithm}[tb]
    \SetAlgoLined
    \DontPrintSemicolon
    \caption{Multi-round recordings simulation.}
    \label{alg:multi-round}
    \SetAlgoVlined
    \SetKwInOut{Input}{Input}
    \SetKw{In}{in}
    \Input{{ $\mathcal{Y}$ \tcp*{\scriptsize{Speaker lists set in WSJ0-mix}}}\\
           { $N_{spk}$ \tcp*{\scriptsize{number of speakers per mixture}}}\\
           { $k_{min},k_{max}$ \tcp*{\scriptsize{Min\&Max number of rounds per mixture}}}
           { $\beta$ \tcp*{\scriptsize{random shift range}}}
           { $R$ \tcp*{\scriptsize{SRN range}}}

}
    \SetKwInOut{Output}{Output}
    \Output{$\mathbb{O}\leftarrow\{o\}$ \tcp*{\scriptsize{Simulated list of mixtures }}}
    \BlankLine
    
    \ForAll{$Y$ $\in$ $\mathcal{Y}$}{
         $o \leftarrow \phi$ \tcp*{initial mixture signal}
         $t \leftarrow 0$ \tcp*{beginning of one mixture}
         \For{$k$ $\in$ $[k_{min},k_{max}]$}{
         \For{$y$ $\in$ $Y$}{  \tcp*{each speaker in one mixture}
            Sample one audio $s$ towards spkear $y$ \\
            Sample SNR $r$ from the given range $R$ \\
            $ s = s\times10^{\frac{r}{20}}$\tcp*{scale with SNR}
            $o[t:]$.add($s$) \tcp*{extend the mixture around the end}
            $t = \text{length}($o$)$ \\
            $t = t + \text{random}(-\beta,+\beta)$
                 
         }
    }
     $\mathbb{O}.\text{append}(o)$}
\end{algorithm}

\subsection{Model details}
For the inference module, we used self-attention based encoder-decoder architecture to predict several possible speakers. For both the encoder and decoder, we used one encoder blocks with 512 attention units containing
eight heads ($M = 1, d_{model} = 512, H = 8$). The size of dimension used in key and value is 64 ($d_k = 64, d_v = 64$). We used 2048 internal units ($d_{ff} = 2048$) in a position-wise feed-forward layer. And, we used the Adam optimizer with the learning rate decayed by a factor of $2\times10^{-1}$ after every 20 epochs. We tested several different configuration in the model architecture, we found that the large number of layers (above 4) resulted in unconvergent training. And the configuration with $M = 2$ shows similar results with $M = 1$.

Different from the original transformer model, we did not feed the output embeddings offset by one position to the next step in decoder. Instead, position $i$ is embedded with a linear layer to $\mathbf{j}_i$ (as shown in Eq.~(\ref{eq:pos})) to serve as input at each step. This is to ensure the decoding process can be done without knowing the order of the true speakers, and the order will be decided after the following extraction module by choosing the best permutation with the $\mathcal{L}_r $. 

For the extraction module, we used the original configure from Conv-TasNet~\cite{luo2018tasnet} with $ N = 256, L = 20, B = 256, H = 512, P =3, X = 8, R = 4$. Also, we noticed the update of the base model in extraction could further improve the performance like the same tendency in~\cite{luo2019conv,luo2019dual}. In this paper, we mainly focus the relative performance over the original TasNet.

For the training strategy, we set a large ratio $\alpha$ in Eq. (\ref{eq:loss}) to balance the $\mathcal{L}_c$ and $\mathcal{L}_r$, which get a large difference in their ranges. To be specific, with training continues, the cross-entropy criterion 
$\mathcal{L}_c$ tends to a small positive number close to zero, while the non-probabilistic $\mathcal{L}_r$ changes from positive to almost -20 because of the negative SI-SNR loss definition. Therefore, we set $\alpha = 50$ to keep a reasonable balance between these two factors. Besides, in practice, we found that the extraction module takes much more time to converge than the speaker inference module. To avoid the overfitting, the speaker inference module is early-stopped based on the $\mathcal{L}_c$ in validation set, which the extraction module will continue until converged.
\vspace{-0.2cm}
\begin{figure}[tb]
  \centering
  \includegraphics[width=8cm]{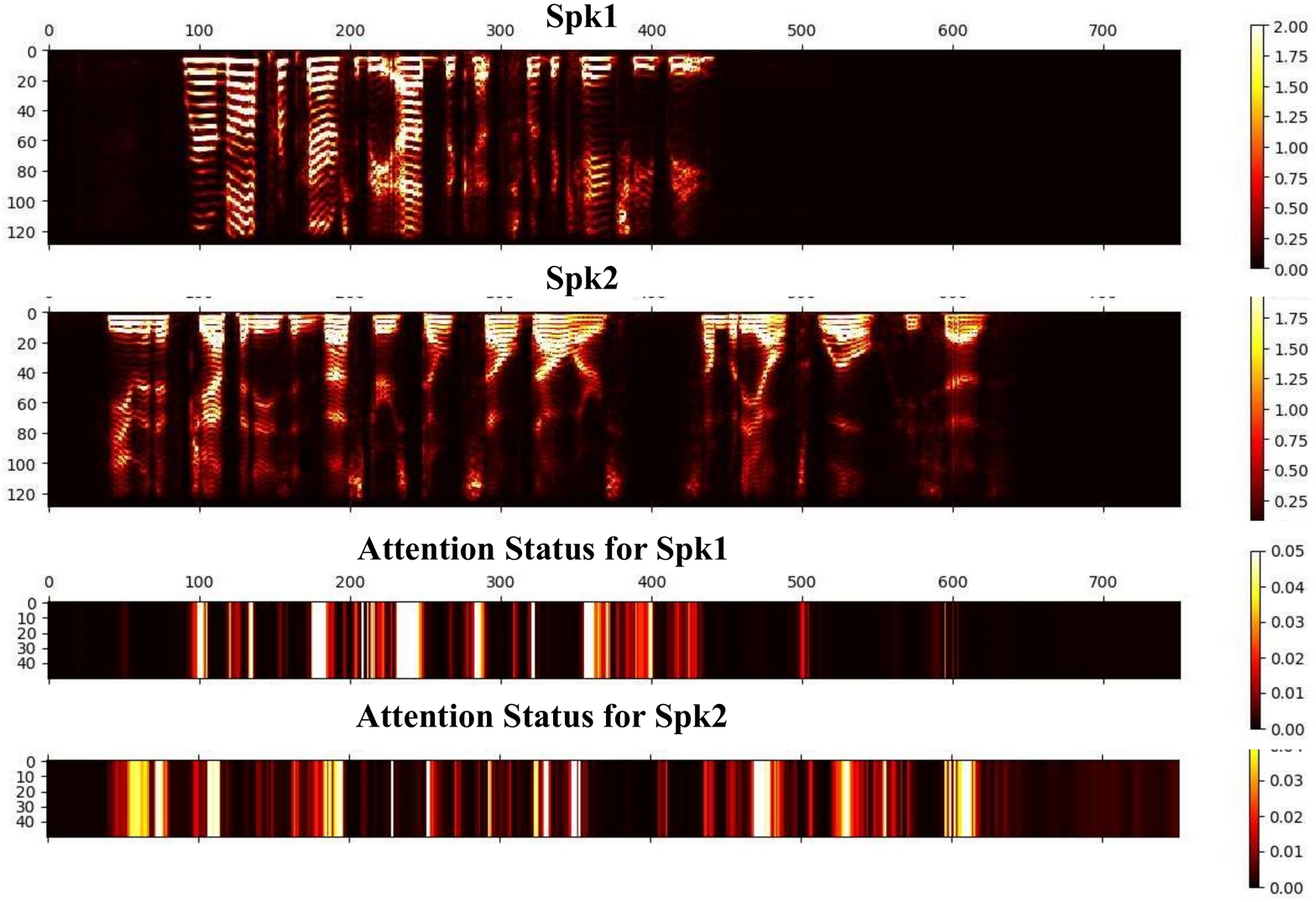}
  \caption{Visualization of one sample of the learned attention status in speaker inference module for overlapped speech in WSJ0-2mix.}
  \label{fig:attention}
\end{figure}
\begin{figure}[tb]
  \centering
  \includegraphics[width=8cm]{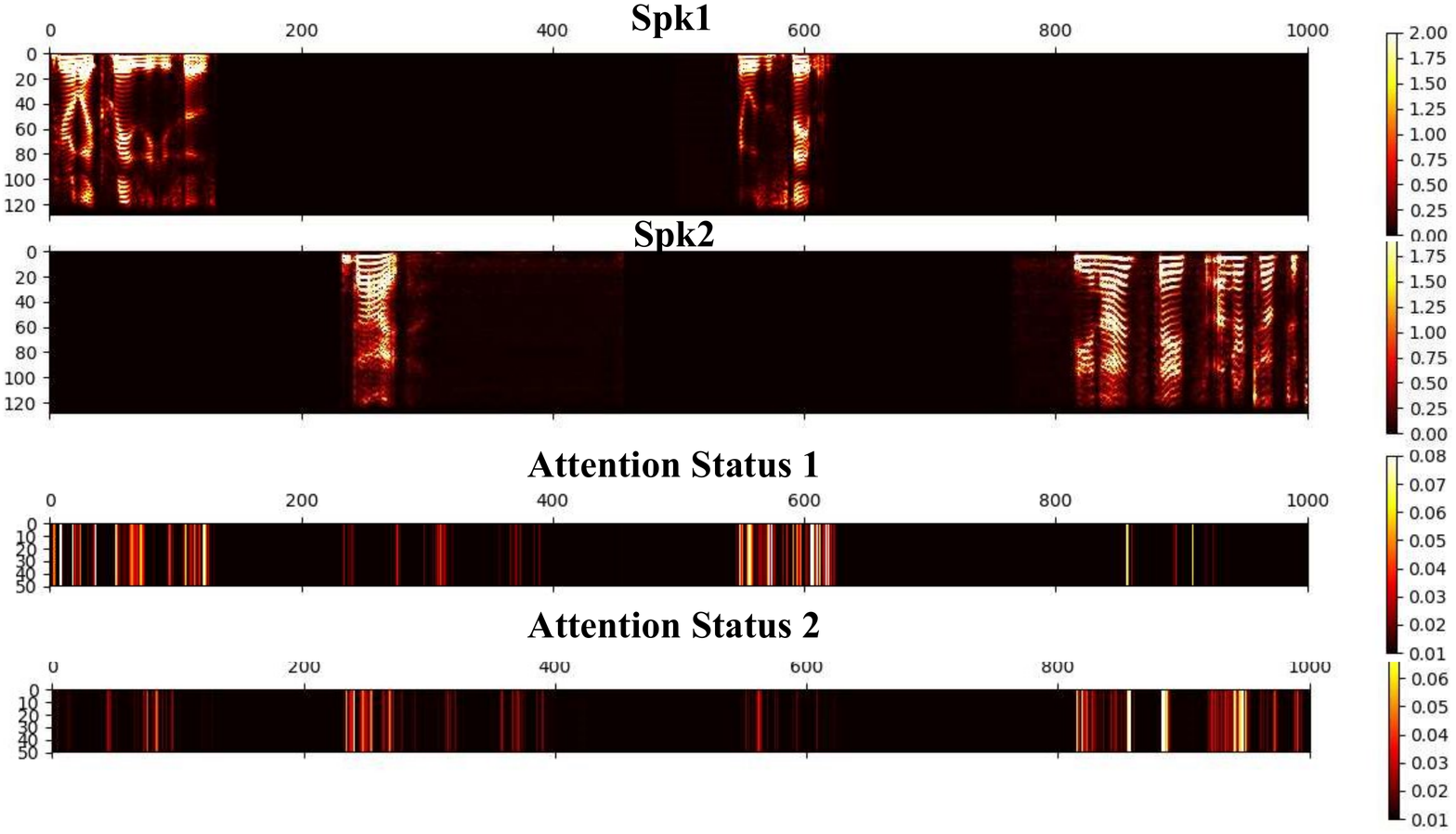}
  \caption{Visualization of one sample of the learned attention status in 2 rounds of utterances.}
 \vspace{-0.1cm}
  \label{fig:attention-2rounds}
\end{figure}
\subsection{Simulation of WSJ0-mix multi-round recordings}
For the multi-round mixtures mentioned in Section~\ref{sec:multi-round}, we simulated them by Algorithm~\ref{alg:multi-round}. The algorithm is to simply simulate the natural conversations with several parts of overlapped part. 

\subsection{Attention status}
Attention mechanisms have become an integral part of compelling sequence modeling and transduction models in various tasks, allowing modeling of dependencies without regard to their distance in the input or output sequences~\cite{vaswani2017attention,bahdanau2014neural,kim2017structured}. 
For the speech related tasks, the vocal characteristics from one specific speaker stay stable in a short segment and a long conversation. Based on these, we use the self-attention based model in our inference part to utilize the relation between different frames from the same speaker. Therefore, the attention status could be used to check the specific process to find the possible speakers. As shown in Figure~\ref{fig:attention}, we visualized one example from WSJ0-2mix test set about
the real spectrograms of the two speakers and the corresponding attention status towards them. The attention status is from the multi-head self-attention block in the decoder, and we added the weights from each head to form the attention status $\in \mathbb{R}^{1 \times \tilde{T}} $. 

As we expect, the attention status shows significant consistency with the real spectrogram. In particular, the attention tends to focus on the frame with larger difference. This is to say, if one speaker gets dominant in some frames, then the attention of this one tends to place emphasis on these dominant frames. Similarly, the attention from multi-round mixture also shows the consistency for one speaker in the whole audio, which could be taken as the implicit speech activity outputted by speaker diarization task.  


\end{appendix}
\end{document}